\documentclass[preprintnumbers, floatfix, preprintnumbers, letterpaper, superscriptaddress,nofootinbib]{revtex4}
\usepackage{graphicx}
\usepackage{microtype}
\usepackage{amsmath}
\usepackage{amssymb}
\usepackage{subfigure}
\usepackage{hyperref}
\usepackage{url}
\usepackage{xcolor}
\usepackage{color}
\usepackage{mathrsfs}
\usepackage{calrsfs}
\usepackage{amsfonts}
\usepackage{eufrak}
\usepackage{tabularx}
\usepackage{eucal}
\usepackage{latexsym}
\usepackage{ragged2e}
\usepackage{epsfig}
\usepackage{textcomp}

\usepackage{caption}
\DeclareCaptionJustification{justified}{\leftskip=0pt \rightskip=0pt \parfillskip=0pt plus 1fil}
\definecolor{vividviolet}{rgb}{0.62, 0.0, 1.0}
\definecolor{amaranth}{rgb}{0.9, 0.17, 0.31}
\definecolor{palatinateblue}{rgb}{0.15, 0.23, 0.89}
\definecolor{brightpink}{rgb}{1.0, 0.0, 0.5}
\definecolor{cornflowerblue}{rgb}{0.39, 0.58, 0.93}
\definecolor{deepcarminepink}{rgb}{0.94, 0.19, 0.22}
\definecolor{radicalred}{rgb}{1.0, 0.21, 0.37}

\hypersetup{ linktoc=all,
	colorlinks, linkcolor={palatinateblue},
	citecolor={brightpink}, urlcolor={amaranth}
}

\graphicspath{{Images/}}

\graphicspath{{Images/}}

\makeatletter
\def\@fnsymbol#1{\ensuremath{\ifcase#1\or \ddagger \or  $\textleaf$  \or \dagger
		\else\@ctrerr\fi}}%

\makeatother

\def\sideremark#1{\ifvmode\leavevmode\fi\vadjust{\vbox to0pt{\vss
			\hbox to 0pt{\hskip\hsize\hskip1em
				\vbox{\hsize1.3cm\tiny\raggedright\pretolerance10000
					\noindent #1\hfill}\hss}\vbox to8pt{\vfil}\vss}}}%
\def\beq{\begin{equation}}
\def\eeq{\end{equation}}

\begin{document}

\title{A Critical Assessment of Black Hole Solutions With a Linear Term \\in Their Redshift
Function}

\author{Daniele \surname{Gregoris}}
\email{danielegregoris@libero.it}
\affiliation{School of Science, Jiangsu University of Science and Technology, Zhenjiang 212003, China}

\author{Yen Chin \surname{Ong}}
\email{ycong@yzu.edu.cn}
\affiliation{Center for Gravitation and Cosmology, College of Physical Science and Technology, Yangzhou University, \\180 Siwangting Road, Yangzhou City, Jiangsu Province  225002, China}
\affiliation{School of Aeronautics and Astronautics, Shanghai Jiao Tong University, Shanghai 200240, China}

\author{Bin \surname{Wang}}
\email{wangb@yzu.edu.cn}
\affiliation{Center for Gravitation and Cosmology, College of Physical Science and Technology, Yangzhou University, \\180 Siwangting Road, Yangzhou City, Jiangsu Province  225002, China}
\affiliation{School of Aeronautics and Astronautics, Shanghai Jiao Tong University, Shanghai 200240, China}

\begin{abstract}
Different theories of gravity can admit the same black hole solution, but the parameters usually have different physical interpretations. In this work we study in depth the linear term $\beta r$ in the redshift function of black holes, which arises in conformal gravity,  de Rham-Gabadadze-Tolley (dRGT) massive gravity, $f(R)$ gravity (as approximate solution) and general relativity. Geometrically we quantify the parameter $\beta$ in terms of the curvature invariants. 
Astrophysically we found that $\beta$ can be expressed in terms of the cosmological constant, the photon orbit radius and the innermost stable circular orbit (ISCO) radius. The metric degeneracy can be broken once black hole thermodynamics is taken into account. Notably, we show that under Hawking evaporation, different physical theories with the same black hole solution (at the level of the metric) can lead to black hole remnants with different values of their physical masses with direct consequences on their viability as dark matter candidates. In particular, the mass of the graviton in massive gravity can be expressed in terms of the cosmological constant and of the formation epoch of the remnant. Furthermore the upper bound of remnant mass can be estimated to be around $0.5 \times 10^{27}$ kg. 
\end{abstract} 

\maketitle

\section{Introduction}

The detection of gravitational wave signals by the LIGO collaboration \cite{ligo1}, and the measurements of the size of the shadow of the black hole Sagittarius A* by the Event Horizon Telescope \cite{eth1,eth2}, have opened new incredible opportunities for testing gravitational theories in their strong field regime. For example restricting our attention to the latter, predictions about the shadow of a Kerr-Newman black hole with \cite{shadow2} or without \cite{shadow1} a cosmological constant, or placed inside an expanding universe \cite{shadow3} have been formulated. From a more fundamental point of view, the observational features of the black hole shadow may constitute a valuable tool for assessing critically the very core foundation of gravitational theories like the no-hair theorem \cite{shadow4,shadow5}, braneworld \cite{shadow6}, and Gauss-Bonnet corrections \cite{shadow7}, and may also be used as indirect evidence of plasma \cite{shadow8}, or of cosmic fluid like dark matter in the surrounding of the black hole \cite{shadow9}. However, testing the Kerr hypothesis does not seem an easy task because some other parameters, like scalar fields or an electric charge, may mimic the effects of the black hole spin \cite{bambis1,bambis2,bambis3}.
More seriously, different gravitational theories can be consistent with the same shadow properties because these theoretical studies rely on the solution of null geodesics about the black hole which depend only on the metric tensor and of its derivatives. In fact, the same spacetime manifold can arise in many different gravitational theories: see e.g. \cite{kerr} for the Kerr black hole. The same problem arises when inferring the black hole properties from the measurements of the size of its accretion disk which is determined by solving the timelike geodesics motion \cite{novikov1,novikov2}. Despite such degeneracies, parameters in the metric of a certain theory might have a different \emph{physical} interpretation in another theory. As a consequence, some physical processes (e.g. thermodynamics) may, in principle, allow us to distinguish the underlying theories.

A static and spherically symmetric black hole solution with a linearly increasing term (as function of the coordinate radial distance) in the redshift function is an example of  degeneracy between (at least\footnote{Recently a mathematical procedure has been developed for reconstructing the gravitational theory, although not always expressible in a closed analytical form, to which a certain black hole solution corresponds, showing that the number of such theories is \emph{infinite} \cite{grg2021}.}) four different gravitational theories. In fact, the spacetime metric
\beq
\label{metric}
ds^2= g_{ab}  dx^a dx^b={\bf e}_a \cdot {\bf e}_b=  -f(r) dt^2+\frac{dr^2}{f(r)}+r^2 d\theta^2 +r^2 \sin^2 \theta d\phi^2\,, \qquad f(r)=1-\frac{2M}{r}+\beta r -\frac{\Lambda r^2}{3}
\eeq
is a solution in conformal gravity \cite{prlconf,linconf1,linconf3},  de Rham-Gabadadze-Tolley (dRGT) massive gravity \cite{de1}, (approximate) $f(R)$ gravity \cite{ori1,origeo}, and general relativity (in this latter case this spacetime is known under the name of {\it generalized Kiselev black hole}) \cite{kis1}.
In each of these theories, the parameters $M$, $\beta$, and $\Lambda$, which describe the same manifold at a geometrical level, carry different physical meanings. The degeneracy of the black hole metric (\ref{metric}) between the conformal and the massive gravity theories has been already addressed in \cite{epl}, but the comparison with the $f(R)$ gravity and general relativity frameworks has not been explored so far. The parameter $\beta$, which makes this solution different from the Schwarzschild-(anti-)de Sitter one, leads to a modified Newtonian dynamics (MOND) \cite{mond} which has been investigated in regard  of deviations in the trajectories of massive bodies in the solar system which would follow from an entropic force  \cite{bhl3},  of possible phase transitions \cite{bhl4}, and as a solution of the galaxy rotation curves problem \cite{linconf1}.  Studies of the quasi-normal frequencies of vibrations of this black hole \cite{qnm1,qnm2}, and of their anomalous behavior \cite{ori2}, have showed that the strong cosmic censorship conjecture can be violated for some choices of the black hole parameters \cite{bhl1}.


In this paper, we will critically assess this degeneracy by firstly describing the geometrical meaning of the parameters $M$, $\Lambda$, and  $\beta$ in terms of some  curvature invariant quantities. This will allow us to provide an algorithm for a local measurements of such parameters which is independent of the previously mentioned different physical interpretations. We will focus our attention on the latter parameter and construct an appropriate combination of curvature invariants for quantifying the \lq\lq distance" between our MOND solution and the Schwarzschild-(anti-)de Sitter spacetime. 
Then, we will illustrate the physical importance of the parameter $\beta$ in the modeling of the innermost stable circular orbit (which can be taken roughly to correspond to the location of the accretion disk), and of the photon sphere (which may be used as an approximation of the shadow size). Also in this case there is a degeneracy between the different physical theories because the analysis is based on the study of the geodesic motion which is a purely geometrical concept. We will then break this degeneracy by looking at the existence of a possible remnant of the black hole (\ref{metric}) after evaporation. In fact, although the radius of the remnant would have the same functional dependence on the black hole parameters, the physical mass and entropy of the remnant would differ between the various gravitational models. Physically, this means that the physical mass and entropy of the remnant would depend differently on the location of the accretion disk and shadow of the progenitor. As a consequence \emph{the formation time of the remnant would be different in different physical frameworks}, whose applicability in accounting for dark matter would be directly affected; we will discuss this point in particular comparing and contrasting the claims that the black hole (\ref{metric}) arises in a spacetime filled with massive gravitons or in a spacetime supported by an anisotropic fluid. 

Our paper is organized as follows: in Sect. \ref{secg} we will investigate the curvature structure of the black hole (\ref{metric}); this section is divided into two parts: in \ref{sIIa} we will present some curvature syzygys both in terms of scalar polynomial curvature invariants and of Cartan curvature invariants; in \ref{sIIb} we will propose a local procedure for a  measurement of the black hole parameters via the curvature invariants for taming the teleological nature of black hole spacetimes.
We will continue with the analyses of the location of the accretion disk and of the shadow in Sect. \ref{sIII}, in which we also point out how the parameter $\beta$ may play the same role as the spin of a rotating Kerr black hole. The physical degeneracies explored in these two sections will be broken in Sect. \ref{sIV} by showing that the different physical frameworks for which (\ref{metric}) is a solution predict different characteristics of the black hole remnant after evaporation.  We will also deepen the analysis of the roles of massive gravitons vs. anisotropic fluid in the possibility of invoking remnants as part of the dark matter budget in \ref{sIVa}. Finally we conclude in Sect. \ref{sV} by providing some physical motivations in support of our mathematical results.

\section{Geometrical interpretation of $\beta$: Curvature structure of the black hole spacetime}
\label{secg}

In this section, we will explore the geometrical meaning of the parameter $\beta$ which provides a linear modification to the black hole redshift function beyond the Schwarzschild-(anti-)de Sitter solution. First of all, we will show how this parameter affects some {\it syzygys}, that is, some algebraic constraints between the curvature invariants at any generic spacetime point, and which therefore characterize the spacetime (\ref{metric})  \cite{exact}. Mathematically, these relationships will provide a local and invariant description of the specific black hole solution we are interested in, and this study would be a preliminary step allowing us to isolate the black hole parameters $\beta$, $M$, and $\Lambda$ as appropriate algebraic combinations of curvature invariants. Thus, a local procedure for the measurements of the values of the black hole parameters will be provided  taming the well-known problems of the teleological nature of black hole spacetimes.  We remark that the curvature objects are simply a geometrical property of the  manifold which are fully determined once the metric tensor (\ref{metric}) is provided without the need of knowing the gravitational theory in which it was discovered.

\subsection{Curvature syzygys}
\label{sIIa}

In the study of the algebraic relationships between various curvature invariants characterizing the black hole spacetime (\ref{metric}), we will adopt three different methods, one relying on the so-called {\it scalar polynomial curvature invariants}, a second one based on the {\it Cartan curvature invariants}, and a third one exploiting the {\it Newman-Penrose curvature scalars}. 
We remind the reader that the Cartan curvature invariants are the components of the curvature tensors and of their derivatives computed in the canonical frame. We will also establish some quantitative relationships between the curvature quantities delivered by these different algorithms. As a side result, we will identify two appropriate curvature invariants which can separately detect locally both the black hole and the cosmological horizons through an algebraic equation.

We begin by analyzing the curvature structure of the black hole spacetime of interest in terms of the scalar polynomial curvature invariants. Let $R_{abcd}$,  $R_{ab}$, $C_{abcd}$ and $G_{ab}$ denote the Riemann curvature,  the Ricci curvature, the Weyl curvature, and the Einstein curvature tensors in the coordinate basis, respectively; let also a semicolon denote a covariant derivative. Let us introduce the following set of scalar polynomial curvature invariants\footnote{We can observe that ${\mathcal I}_1$ corresponds to the Kretschmann scalar.} \cite{mac}:
\begin{eqnarray}
	{\mathcal I}_1 &:=&  R_{abcd} R^{abcd}=\frac{8(\Lambda^2 r^6-3\Lambda\beta r^5+3\beta^2 r^4+18 M^2)}{3 r^6} \,,\\
	\label{hord1}
	{\mathcal I}_2 &:=&  R_{abcd;e} R^{abcd;e}= \frac{16(\beta^2 r^4 +45 M^2)(3r -6M -\Lambda r^3 +3\beta r^2)}{3 r^9}\,, \\
	{\mathcal I}_3 &:=&  R_{,a} R^{,a}=\frac{12 \beta^2 (3r -6M -\Lambda r^3 +3\beta r^2)}{r^5}  \,, \\
	{\mathcal I}_4 &:=&  R_{ab} R^{ab}=\frac{2(2\Lambda^2 r^2-6\Lambda\beta r+5\beta^2)}{r^2} \,,\\
	{\mathcal I}_5 &:=&  R_{ab} G^{ab}= \frac{4 (\beta-\Lambda r)(\Lambda r-2\beta)}{r^2} \,, \\
	{\mathcal I}_6 &:=&  G_{ab} G^{ab}= \frac{2(2\Lambda^2 r^2-6 \Lambda \beta r+5\beta^2)}{r^2} \,, \\
	{\mathcal I}_7 &:=&  C_{abcd;e} C^{abcd;e}= \frac{240(3r -6M -\Lambda r^3 +3\beta r^2)}{ r^9}\left(\frac{4 r^4 \beta^2}{45} +M^2 \right)\,, \\
	\label{eqR}
	R &=& 4\Lambda - \frac{6 \beta}{r}  \,.
\end{eqnarray}
First of all, we should observe that 
\beq
{\mathcal I}_7 - {\mathcal I}_2 =\frac{48 \beta^2 f(r)}{r^4} 
\eeq
can be zero in regions other than the horizon if and only if $\beta=0$, i.e. the MOND effects are suppressed.
By direct inspection, we see that the following  syzygy holds:
\beq
\sqrt{\frac{(9 {\mathcal I}_2 -4 {\mathcal I}_3)(4{\mathcal I}_4 -R^2)}{5 {\mathcal I}_3}}-\sqrt{\frac{6({\mathcal I}_1-{\mathcal I}_5 - {\mathcal I}_6)-R^2}{2}}   =0\,.
\eeq
Let us now introduce the following null coframe:
\beq
\label{coframe}
l^a=\frac{1}{\sqrt{2}}\left(\sqrt{f(r)}dt -\frac{dr}{\sqrt{f(r)}} \right)\,, \qquad n^a=\frac{1}{\sqrt{2}}\left(\sqrt{f(r)}dt +\frac{dr}{\sqrt{f(r)}} \right)\,, \qquad m^a=\frac{1}{\sqrt{2}}\left(r d\theta +i \,r \sin \theta d\phi \right)\,,
\eeq
where $i^2=-1$, in terms of which the spacetime metric (\ref{metric}) reads as
\beq
ds^2=-2 l_{(a}n_{b)}+2 m_{(a}\bar m_{b)}\,,
\eeq
where an overbar denotes a complex conjugation, and round parentheses stand for symmetrization. We note that the relationships $l_a l^a=n_a n^a=m_a m^a =\bar m_a \bar m^a=0$ and $ -l_a n^a=1=m_a \bar m^a$ hold. In the coframe (\ref{coframe}) we obtain the following results for the non-zero Newman-Penrose curvature scalars\footnote{In the following equations we have denoted with $\Lambda_{\rm NP} $ the Newman-Penrose curvature scalar which is usually denoted by $\Lambda$  for avoiding confusion  with the cosmological constant parameter.} \cite{NPC}:
\begin{eqnarray}
	\Lambda_{\rm NP} &=&\frac{\Lambda}{6} -\frac{\beta}{4r} \,, \\
	\Phi_{11} &:=&\frac{1}{4}S_{ab}(n^a l^b +m^a \bar m^b)=-\frac{\beta}{4r}    \,, \\
	\label{psi2}
	\Psi_2&:=& -C_{abcd} n^a m^b l^c \bar m^d= -\frac{M}{r^3} \,, 
\end{eqnarray}
in which we have introduced as auxiliary quantity the tracefree tensor
\beq
S_{ab}:= R_{ab}-g_{ab}\frac{R}{4}\,.
\eeq
Thus, the 1-forms (\ref{coframe})  constitute the canonical coframe for the metric tensor (\ref{metric}). By this we mean that while the metric tensor is invariant under the transformations constituting the Lorentz group, i.e. spin-boost and null rotations governed by
\beq
l^a \to f_1^2 l^a\,, \qquad n^a \to \frac{1}{f_1^2} n^a \,,\qquad m^a \to e^{2 i f_2}m^a  \,,
\eeq
and
\beq
l^a \to l^a\,, \qquad n \to n^a +f_3 m^a +\bar f_3 \bar m^a +f_3 \bar f_3 l^a\,, \qquad m^a \to m^a + \bar f_3 l^a  \,,
\eeq
respectively, where $f_1=f_1(t,r,\theta,\phi)$, $f_2=f_2(t,r,\theta,\phi)$,  and $f_3=f_3(t,r,\theta,\phi)$ are arbitrary functions of the manifold coordinates,  the components of the curvature tensors (and thus the Newman-Penrose scalars) may change. However, once we have fixed the null coframe (\ref{coframe}),  the curvature tensor has been reduced to its canonical form for a spacetime whose symmetry group is $SO(1,3)$ (see appendix B in \cite{advgr}). We discover that the following two syzygys between scalar polynomial curvature invariants and Newman-Penrose scalars hold:
\begin{eqnarray}
	\label{ss1}
	&& \frac{\sqrt{2[6({\mathcal I}_1 -{\mathcal I}_5-{\mathcal I}_6)-R^2]}}{24}	\,=\, \frac{M}{r^3}\,=\, -\Psi_2 \\
	\label{ss2}
	&& \frac{\sqrt{4 {\mathcal I}_4 -R^2}}{2}	\,=\, \frac{\beta}{r}\,=\, -4 \Phi_{11} \,,
\end{eqnarray}
which can be combined into:
\beq
\label{distance}
3 \sqrt{\frac{(4 {\mathcal I}_4 -R^2)^3}{2[6({\mathcal I}_1 -{\mathcal I}_5-{\mathcal I}_6)-R^2]}} =\frac{64 \Phi_{11}^3}{\Psi_2} \equiv \frac{\beta^3}{M}\,.
\eeq
This latter quantity may be interpreted as an invariant measure of the {\it distance} between the black hole solution (\ref{metric}) and the Schwarzschild-(anti-)de Sitter spacetime, because it vanishes if and only if the $\beta$ parameter is zero. Thus, this specific combination of the scalar polynomial curvature invariants (or equivalently of the Newman-Penrose scalars) plays the same role of the {\it ``Kerness''} and {\it ``massiveness''}  proposed respectively in \cite{lake} and in \cite{mio3} for quantifying respectively the {\it distance} between  axisymmetric Kerr black holes and more realistic distorted astrophysical black holes, and analogously the distance between Banados-Teitelboim-Zanelli black holes arising in general relativity and in massive gravity. Although we have shown that for the black hole (\ref{metric}) such a distance could be constructed equivalently in terms of either the scalar polynomial curvature invariants or using the Newman-Penrose scalars,  we will soon explain how these two different paths exhibit different computational applicabilities when trying to cast the black hole parameters as appropriate combinations of curvature quantities. In fact, the former method would require us to compute second degree quantities in the  derivatives of the curvature components, while the latter procedure would be based  on first degree curvature quantities,  making it computationally more efficient.

We remark that the Newman-Penrose scalar $\Phi_{11}$ is non-zero even though the spacetime (\ref{metric}) may be regarded as a vacuum solution for the $f(R)$ theory, for the Weyl gravity theory, and for the massive gravity theory: this is indeed in agreement with the claim that a modification of the gravity sector behaves effectively as an extra scalar field (i.e. an extra matter component entering the stress-energy tensor) in the general relativity picture \cite{sergei}. The result obtained for $\Phi_{11}$ is consistent also with the observation that the $f(R)$ gravity model in which (\ref{metric}) was discovered
 reduces to general relativity on large distances at which $\Phi_{11} \to 0$.

With respect to the coframe (\ref{coframe}),  we can compute the following interesting Cartan curvature invariants \cite{cartanref}:
\begin{eqnarray}
	\label{cartan1}
	{\mathcal J}_1 &:=&  C_{\theta r \phi t} =\frac{M}{r^3} \,,\\
	\label{hord2}
	{\mathcal J}_2 &:=& {\bf e}_r (C_{\phi\theta\phi\theta})= {\bf e}_r (C_{rtrt})= \frac{3 M \sqrt{2 f(r)}}{2 r^4} \,, \\
	\label{cartan3}
	{\mathcal J}_3 &:=&  R_{\theta \phi \phi \theta } =\frac{\Lambda}{3}-\frac{\beta}{r}+\frac{M}{2 r^3} \,,\\
	\label{cartan4}
	{\mathcal J}_4 &:=&  R_{rttr} =\frac{\Lambda}{3}+\frac{M}{2 r^3} \,,\\
	\label{cartan5}
	{\mathcal J}_5 &:=&  R_{tr}=\frac{\beta}{r} -\Lambda \,.
\end{eqnarray}
Thus, comparing (\ref{cartan1}) against (\ref{ss1}), and ${\mathcal J}_4 -{\mathcal J}_3$ versus (\ref{ss2}), two new syzygys relating scalar polynomial curvature invariants and Cartan invariants are discovered. As done before, we can re-formulate the measure of the {\it distance}  between the black hole (\ref{metric}) and the Schwarzschild-(anti-)de Sitter spacetime as:
\beq
\label{distance2}
\frac{({\mathcal J}_4 -{\mathcal J}_3)^3}{{\mathcal J}_1} \equiv \frac{\beta^3}{M}\,,
\eeq
and again, even though there are no conceptual differences between these various routes which deliver the same final result, we will clarify some important different applicabilities of these different methods  in light of practical applications in numerical relativity.

The two invariant curvature quantities (\ref{hord1}) and (\ref{hord2}) separately detect the horizon(s) of the black hole (\ref{metric}) because they vanish in their correspondence and only there (they do not provide any false positives). From the conceptual point of view, the meaning of this finding is that it is not necessary to know the entire evolution of the black hole spacetime, e.g. its behavior at null spatial infinity, for locating its horizon(s) which can therefore be observed by experimental devices of finite sizes, taming its teleological nature pointed out in \cite{BIN1,BIN2}. Therefore, we have explicitly shown that the horizon constitutes a local property of the manifold (in agreement with the core principles  of any relativistic field theory), and that a curvature invariant can be constructed for its detection once the geometrical symmetries of the spacetime are known regardless the gravitational theory behind it: this is consistent with the {\it geometric horizon conjecture} \cite{shoom,sho1,app1,app3,app4,app5,mio1,mio2,mio3,turco}. On the other hand, our results are important also from the practical point of view in light of the so-called {\it excision}  technique in numerical relativity: the black hole horizon constitutes a causal boundary separating the evolutions of  phenomena occurring outside it from what it may happen inside; thus, the spacetime region delimited by the horizon must be removed (or excised) when performing numerical simulations of the evolution of a black hole. Standard excision techniques rely on the solution of the Raychaudhuri equation which accounts for the propagation of a bundle of light beams exploiting the change of their focusing properties taking place on the horizon \cite{num1,num2,num3}. This method requires integrating a differential equation with the need of knowing information on more than one spacetime point. However, according to our two proposals  for finding this region it would be necessary only to search for the roots of an algebraic equation which imposes either the curvature quantity (\ref{hord1}) or (\ref{hord2}) to vanish making it computationally more advantageous. Moreover, the latter choice based on a Cartan invariant seems to be even more convenient than the former because it is based on a first degree quantity in the curvature rather than on a second degree one. Furthermore, in the latter approach the position of the horizon is found by looking for the zeroes of the frame derivative of the Weyl curvature; this can be related to searching for local extrema in the radial evolution of the tidal force generated by the black hole, as considered in a number of specific solutions \cite{tidal1,tidal2}, but our approach is grounded on the {\it geometric horizon conjecture} rather than on the solution of the geodesic deviation equation.

\subsection{A local measurement  of the black hole parameters}
\label{sIIb}

So far we have explained why some appropriate curvature quantities are computationally useful for detecting the horizon of the black hole (\ref{metric}). However, numerical simulations of black hole systems require to know also the values of the parameters characterizing the particular solution in hand,  which are $M$, $\Lambda$, and $\beta$ in our case. Such values are usually estimated from the size of the area of the horizon  \cite{rez}; since it is necessary to evaluate an integral for computing this area, a procedure has been proposed in \cite{lake}  for isolating the black hole parameters for the Kerr black hole, like its mass and spin, as appropriate algebraic combinations of some scalar polynomial curvature invariants. It is conceivable, although not proved yet, that similar results may hold for any black hole solution. As for the analysis dealing with the detection of the horizon, this finding would confirm that the physical quantities characterizing the black hole are indeed local, but it would also provide a more convenient procedure for estimating them in numerical computations. Thus, the seminal result of \cite{lake} has been extended to the case of the Kerr-Newman-NUT-(Anti)-de Sitter black hole in \cite{mio1} by using both scalar polynomial curvature invariants and Cartan curvature invariants separately, and to the case of the massive Banados-Teitelboim-Zanelli  black hole in \cite{mio3} by using a mixture of these two types of curvature quantities.

However, these results are extremely non-trivial because a constructive procedure which can deliver the black hole parameters in terms of curvature invariants does not exist in literature. This means that the physical quantities must be isolated by hands case by case looking at the specific black hole metric under investigation through a trial-and-error procedure. Despite the difficulty of this task, we were able to relate the parameters of the black hole solution (\ref{metric}) to appropriate combinations of some curvature invariants by exploiting the results about its curvature structure exhibited previously. Explicitly, from the scalar polynomial curvature invariants we can get:
\begin{eqnarray}
	\label{betaiso}
	\beta &=& (4{\mathcal I}_4  -R^2)  \sqrt{\frac{3 }{(4{\mathcal I}_4  -R^2)[\sqrt{2[6({\mathcal I}_1 -{\mathcal I}_5-{\mathcal I}_6)-R^2]}-3\sqrt{4{\mathcal I}_4  -R^2}+R]-{\mathcal I}_2+{\mathcal I}_7}}     \,, \\
	\Lambda &=&\frac{R +3\sqrt{4{\mathcal I}_4  -R^2}}{4}\,,
\end{eqnarray}
and then the mass $M$ can be obtained equivalently from (\ref{distance}) or from (\ref{distance2}) using either scalar polynomial curvature invariants, or Newman-Penrose scalars, or Cartan curvature invariants. Moreover, using only Newman-Penrose scalars we also obtain 
\beq
\Lambda= 6(\Lambda_{\rm NP}-\Phi_{11})\,,
\eeq
while using only Cartan curvature invariants we can isolate the cosmological constant as
\beq
\Lambda=  {\mathcal J}_4 -{\mathcal J}_3 -{\mathcal J}_5 \,.
\eeq
While there are no conceptual differences in these alternative but equivalent algorithms for measuring locally the black hole parameters, they exhibit different applicability when practical computations are performed. In fact, we have shown that the method relying on the Newman-Penrose scalars or on the Cartan curvature invariants is preferable than the one using scalar polynomial curvature invariants, as they are less expensive computationally, being fully based only on first degree quantities. We remark also that Cartan invariants are foliation-independent \cite{cartanref}. It should be noted that as for the discussion about locating the horizon by using an appropriate scalar polynomial curvature invariant or a Cartan invariant, we have used here only the geometrical properties of the black hole solution (\ref{metric}) and not its correspondence to any specific gravitational framework.

\section{Physical interpretation of $\beta$: accretion disk and shadow}
\label{sIII}

In this section, we will present the physical relevance of the parameter $\beta$ by claiming that its non-zero value can mimic the spin of a Kerr black hole in accounting for the size of the accretion disk  of black holes. The idea that some parameters of static black hole solutions can be astrophysically indistinguishable from the spin of a rotating black hole  in reference to this dataset has already been pioneered for nonsingular black holes in nonlinear electrodynamics \cite{mr31},  charged stringy black holes \cite{mr35}, deformed black holes \cite{bambi2}, and black holes surrounded by dark matter \cite{bambi3}. In the case of our paper, we are  in an even more extreme situation since we cannot provide an unambiguous physical interpretation to the parameter $\beta$ because these analyses rely on the study of the geodesic motion which depends on the metric tensor and not the gravitational theory formulation. To our knowledge, this degeneracy has never been pointed out, despite the geodesic motion in the spacetime (\ref{metric}) has been extensively studied in the literature \cite{bhl2,origeo,ineq2,nodm1,nodm2,nodm3,geo1,geo2,geo3,geo4,geo5,geo6,geo7,geo8,geo9,geo10,geo11,geo12,geo13,geo14,geo15}.

To begin with, the Novikov-Thorne model, which is widely adopted for the description of the formation of an accretion disk around a black hole, states that the particles constituting the accretion disk follows nearly geodesics paths well approximated by the innermost stable circular orbit (ISCO) falling into the central massive object due to the effects of its gravitational field \cite{novikov1,novikov2}. This makes the ISCOs especially  relevant in astrophysics.
In the spacetime of a static black hole, the ISCO is located at the radius $r=r_{\rm ISCO}$ such that:
\beq
V_{\rm eff}(r_{\rm ISCO})={\mathcal E}\,, \qquad V'_{\rm eff}(r_{\rm ISCO})=0\,, \qquad V''_{\rm eff}(r_{\rm ISCO})= 0\,, 
\eeq
where ${\mathcal E}=E/m$ is the specific energy of the test particle with mass $m$ and energy $E$ (which is conserved during the geodesic motion because the spacetime is static), and the effective gravitational potential should be written in terms of the black hole redshift function and of the specific angular momentum of the test particle $L$ as $V_{\rm eff}=f(r)(1+L^2/r^2)$ \cite{schutz}. For the black hole (\ref{metric}) these three conditions have been used for isolating the cosmological constant as \cite[Eq.(3.5)]{bhl2}:
\beq
\label{Laccr}
\tilde \Lambda= 3\frac{\tilde \beta^2 \tilde r^4_{\rm ISCO} +3 \tilde \beta r_{\rm ISCO}^2(\tilde r_{\rm ISCO}-4)+2(\tilde r_{\rm ISCO}-6)}{r_{\rm ISCO}^3(3\tilde \beta \tilde r_{\rm ISCO}^2 +8 \tilde r_{\rm ISCO} -30)}\,,
\eeq
where 
\beq
\label{rescaling}
\tilde r := \frac{r}{M}\,, \qquad \tilde \Lambda :=\Lambda M^2 \,, \qquad \tilde \beta:=\beta M\,.
\eeq
Using this result and that for the ISCO in a Kerr spacetime \cite{bardeen},
\beq
\tilde r_{\rm ISCO}=3+Z_2 \pm \sqrt{(3-Z_1)(3+Z_1+2Z_2)}\,, \qquad Z_1= 1+((1+a)^{\frac{1}{3}} + (1-a)^{\frac{1}{3}} ) (1 -a^2)^{\frac{1}{3}}  \,, \qquad
Z_2 = \sqrt{3a^2 +Z_1^2}\,,
\eeq
where the plus/minus signs are to be applied for test particles following retrograde/prograde orbits,  we can express the degeneracy between the parameters ($\beta$, $\Lambda$) and the spin parameter $a$ as:
\beq
\label{betaisco}
\tilde \beta=\frac{3\tilde \Lambda \tilde r_{\rm ISCO}^3-9 \tilde r_{\rm ISCO}+36 \pm \sqrt{9 \Lambda^2 \tilde r_{\rm ISCO}^6+42 \Lambda \tilde r_{\rm ISCO}^4-144 \tilde \Lambda \tilde r_{\rm ISCO}^3+81 \tilde r_{\rm ISCO}^2-720 \tilde r_{\rm ISCO}+1728}}{6 \tilde r_{\rm ISCO}^2}\,.
\eeq
We remark that (\ref{Laccr}) is consistent with the asymptotically flat limiting case $\tilde r_{\rm ISCO}=6$ when $\tilde \Lambda =\tilde \beta=0$.
Furthermore, we can approximate the size of the shadow with the radius $r_p$ of the photon sphere which, for a static and spherically symmetric black hole, is given by \cite{shadow8,shadow9}:
\beq
0=\frac{d}{dr}\left(\frac{r^2}{f(r)} \right)\Big |_{r=r_p}\,.
\eeq
Therefore, for the case in (\ref{metric}) we obtain
\beq
\label{shadow}
\tilde r_p=\frac{-1\pm\sqrt{6 \tilde \beta+1}}{\tilde \beta}\,,
\eeq
where the $\pm$ signs hold for positive/negative $\beta$ respectively, and in the latter case as long the mass is such that $M<-1/(6\beta)$. It should be noted that the cosmological constant does not enter this expression, and that once again this result is not sensitive to the gravitational theory adopted for interpreting the black hole metric (\ref{metric}). For small positive $\beta$ we can approximate 
\beq
\label{psmall}
\tilde r_p \approx 3 \left(1-\frac{3\tilde \beta}{2} \right)\,,
\eeq
which shows that the photon sphere is smaller than the one in a Schwarzschild spacetime; this is because for escaping  the photons need to climb a larger potential well. Physically, this may be interpreted as some amount of energy that the photons may lose when scattering against the cosmic fluid (if we adopt the Kiselev interpretation) or for winning the effects of massive gravitons. 
Should the size of the shadow be known from astrophysical observations, we can infer the value of the parameter $\beta$ as
\beq
\tilde \beta= \frac{2(3-\tilde r_p)}{\tilde r^2_p}\,.
\eeq
Implementing this result into the information about the size of the accretion disk (\ref{Laccr}) delivers the following estimate of the rescaled cosmological constant:
\beq
\label{lisco1}
\tilde \Lambda=-\frac{3(2\tilde r_{\rm ISCO}^4 \tilde r_p^2-3\tilde r_{\rm ISCO}^3 \tilde r_p^3-12 \tilde r_{\rm ISCO}^4 \tilde r_p+9\tilde r_{\rm ISCO}^3 \tilde r_p^2+12\tilde r_{\rm ISCO}^2 \tilde r_p^3+\tilde r_{\rm ISCO} \tilde r_p^4+18 \tilde r_{\rm ISCO}^4-36 \tilde r_{\rm ISCO}^2 \tilde r_p^2-6 \tilde r_p^4) }{\tilde r_p^2 \tilde r_{\rm ISCO}^3(3\tilde r_{\rm ISCO}^2 \tilde r_p-4\tilde r_{\rm ISCO} \tilde r_p^2-9\tilde r_{\rm ISCO}^2+15 \tilde r_p^2)}\,.
\eeq
By inserting back the parameter $M$ in (\ref{Laccr}), and eliminating it through the non-rescaled version of (\ref{shadow}), we can obtain:
\beq
\label{lisco}
\Lambda=\Lambda(\beta)= \frac{(3r_{\rm ISCO}^4-6r_{\rm ISCO}^2r_p^2-r_p^4)\beta^2+(9r_{\rm ISCO}^3-12r_{\rm ISCO}^2r_p+r_{\rm ISCO}r_p^2-4r_p^3)\beta+2r_p(r_{\rm ISCO}-2r_p)}{r_{\rm ISCO}^3 [(3r_{\rm ISCO}^2 -5 r_p^2) \beta + 8r_{\rm ISCO}-10r_p]}\,,
\eeq
which will be useful in what follows.

\section{Breaking the degeneracy: mass and entropy of the remnant}
\label{sIV}

The discussion about the curvature structure of the spacetime manifold, and of the properties of its accretion disk and shadow we have carried out in the previous sections are purely geometrical results.
We will now show that the degeneracy between conformal gravity, massive gravity, $f(R)$ gravity, and general relativity, which all admit the black hole solution (\ref{metric}) can be broken by looking at the entropy and at the physical mass of the black hole. We will also propose a scalar-tensor model which can provide as well a black hole solution with a linear term in the redshift. Our main objective is to illuminate the different physical meaning that the black hole parameters $M$, $\beta$, and $\Lambda$ assume in each of these theoretical frameworks.

We recall that conformal gravity is derived from the action
\beq
\label{confS}
S_{\text{conf}}=\int d^4 x \sqrt{-g} C_{\alpha\beta\gamma\delta} C^{\alpha\beta\gamma\delta}\,,
\eeq
in which $C_{\alpha\beta\gamma\delta}$ denotes the Weyl curvature tensor; in this case all the three parameters $M$, $\Lambda$ and $\beta$ do not enter the Lagrangian formulation but appear only as purely mathematical (with no physical meaning) integration constants of the field equations. Indeed in \cite{ineq2} it was claimed that the Modified Newtonian Dynamics (MOND) effects encoded in the linear part of the gravitational potential
\beq
V(r)= -\frac{2 M}{r}+\beta r -\frac{\Lambda r^2}{3}
\eeq
affecting the equations of  geodesic motion  do not have a well-justified physical ground  because the authors interpreted the black hole solution as arising in conformal gravity. Furthermore, assuming (\ref{metric}) to be a solution of (\ref{confS}) the process of black hole evaporation was investigated in \cite{confmass1,confmass2} showing that the black hole physical mass is\footnote{Note that it is well-defined as long as $\Lambda$ is negative if we choose $M$ to be positive.}:
\beq
\label{confmass}
{\mathcal M}_1=-\frac{M\Lambda}{3 \pi}\,.
\eeq

On the other hand, in massive gravity the parameters $M$ and $\Lambda$ are still  mathematical integration constants of the field equations, but instead $\beta$ enters already at the Lagrangian level and it is interpreted to be (related to) the graviton mass $m$: this massive gravity theory is based on the action
\beq
\label{massiveS}
S_{\text{dRGT}}=\frac{1}{16 \pi}\int d^4 x \sqrt{-g}(R+m^2 {\mathcal U}(g, \phi^a))\,,
\eeq
where $R$ is the Ricci scalar and ${\mathcal U}$ is the modified gravitational potential. For example, in \cite{ongmassive} the existence of a black hole remnant after evaporation was interpreted as due to the role played by a massive graviton; it was also shown that the radius, mass and the entropy of the remnant are\footnote{Note that different sign conventions for $\Lambda$ in (\ref{metric}) and in \cite{ongmassive} are adopted.}:
\begin{eqnarray}
\label{mrmassive}
&& R_r=\frac{\beta \pm \sqrt{\beta^2 +\Lambda}}{\Lambda}\,, \qquad M_r=\frac{1}{3}\left(1+\frac{\beta}{2} R_r \right) R_r=\frac{ \beta(\beta \pm \sqrt{\beta^2 +\Lambda} ) -2\Lambda}{6\Lambda^2}( \beta \pm \sqrt{\beta^2 +\Lambda} ) \,, \\
\label{entropymassive}
&& S_r=\pi R^2_r= \frac{\pi ( \beta \pm \sqrt{\beta^2 +\Lambda}  )^2}{\Lambda^2}  \,,
\end{eqnarray}
where $R_r$ is the radius of the configuration. Two solutions for the radius of the remnant exist. It should be noted however that for $\Lambda>0$ the solution with the minus sign should be ignored because a well-defined $R_r>0$ would require $\beta>\sqrt{\beta^2 +\Lambda}$, i.e. $\Lambda<0$. For $\Lambda<0$, the well-posedness of $R_r$ would instead be written as $\beta<\sqrt{\beta^2 +\Lambda}$ (because a multiplication by a negative factor switches the direction of the inequality) which is fulfilled if $\beta<-\sqrt{|\Lambda|}$. Moreover, again taking into account that a multiplication by a negative factor switches the sense of the inequality the  condition $M_r>0$ reads as $\sqrt{\beta^2 +\Lambda}<\frac{2 \Lambda}{\beta}+\beta$. Here, the RHS is positive if $\beta^2<-2\Lambda$; then we would obtain $-3\Lambda \beta^2 <4\Lambda^2$, i.e. $3\beta^2<-4\Lambda$ should hold, which is fulfilled if $-2 \sqrt{|\Lambda|/3}<\beta<-\sqrt{|\Lambda|}$.
About the plus sign: for a positive cosmological constant it is guaranteed that $R_r>0$, while the mass of the remnant is automatically well-defined if $ \beta>0$, while should this parameter be negative, this condition becomes $\sqrt{\beta^2 +\Lambda}<-\left( \frac{2\Lambda}{\beta}+\beta\right)$, which is consistent because $\beta^2 + 2\Lambda>0$, implying
$-3\Lambda\beta^2<4\Lambda^2$ which holds noting that the LHS is negative. On the other hand, 
for a negative cosmological constant we would need to satisfy the condition $\sqrt{\beta^2 +\Lambda}<-\beta$ which can be achieved again if $\beta<-\sqrt{|\Lambda|}$. Then, imposing a positive $M_r$ and repeating the same steps as previously, we obtain $\sqrt{\beta^2 +\Lambda}>-\left( \frac{2\Lambda}{\beta}+\beta\right)$, requiring us to restrict the range $\beta<-\sqrt{2|\Lambda|}$; then we can write $3\beta^2>-4\Lambda$ which is satisfied in our range.
These restrictions are summarized in Table  \ref{table1}. It should be remarked that there is no ambiguity on which sign should be considered in front of the square root in the expression for $R_r$  once the range of the parameters is decided, and that remnants in massive gravity can exist in both asymptotically de Sitter and anti-de Sitter spacetimes (the analysis in \cite{ongmassive,ongmassive2} only assumed one possibility for the sign in (\ref{mrmassive})).

\begin{table}
	\begin{center}
		\begin{tabular}{|c|c|}
			\hline
			$R_r $ &  Restrictions on the parameters   \\
			\hline
			 $R_r=\frac{\beta + \sqrt{\beta^2 +\Lambda}}{\Lambda}$    & $\Lambda>0$   or $\Lambda<0$ and $\beta<-\sqrt{2|\Lambda|}$     \\
			 $R_r=\frac{\beta - \sqrt{\beta^2 +\Lambda}}{\Lambda} $ &  $ \Lambda<0$ and $-2\sqrt{|\Lambda|/3}<\beta<-\sqrt{|\Lambda|}$  \\
			\hline
		\end{tabular}
		\caption{The Table summarizes the ranges of applicability of the two solutions (\ref{mrmassive}) for the radius of the black hole remnant $R_r$ according to the values of the cosmological constant $\Lambda$ and of the parameter $\beta$ requiring that $R_r,M_r>0$ in massive gravity.}
		\label{table1}
	\end{center}
\end{table}

Massive gravity does not belong to the class of scalar-tensor theories \cite{revmassive}, and thus when we will show later that the black hole (\ref{metric}) can arise in the Brans-Dicke framework, it should be considered as a nontrivial correspondence.

Furthermore, the black hole spacetime (\ref{metric}) is a solution also of the (approximate) theory
\beq
\label{model}
S_{f(R)}=\frac{1}{2}\int d^4x \sqrt{-g} \,f(R)\,, \qquad
f(R)=R +\Lambda +\frac{R+\Lambda}{R/R_0 +2/\alpha}\ln \left(\frac{R+\Lambda}{R_c} \right),
\eeq
in which the cosmological constant $\Lambda$ appears already at the level of the Lagrangian. In this formulation $R_c$ is a constant of integration, and $R_0=6 \alpha^2/d^2$ accounts for the two free parameters  $\alpha$ and $d$  whose effects can be  summarized into the effective black hole parameter $\beta=\alpha/d$. The gravity model (\ref{model})  has been proposed in literature because it can pass solar system tests on small length scale regime, i.e. in the limiting case  $r\ll d$, $R\gg\Lambda$, and $R/R_0\gg2/\alpha$, in which the Lagrangian can be approximated as\footnote{We refer to \cite{odint} for explorations of the inflationary and late-time accelerated epochs of logarithmic-corrected gravity.}
\beq
\label{logS}
f(R) \simeq R+R_0 \ln \frac{R}{R_c}\,.
\eeq
On cosmological scales (\ref{model}) reduces to the Einstein-Hilbert Lagrangian $f(R)\simeq R+\Lambda$ adopted in general relativity. Therefore, the parameter $\beta$ enters the Lagrangian as it does in the massive gravity theory and it quantifies again the deviations from general relativity, but is not directly related to the graviton mass. We can compute the physical mass of the black hole (\ref{metric}) interpreted as a solution of the theory (\ref{logS}), i.e. its Misner-Sharp mass which is the conserved charge associated with the timelike Killing Vector Field \cite{misner}, by following the procedure of \cite{faraoni1,faraoni2}:
\beq
{\mathcal M}_2=\left[\frac{df(R)}{2dR}{\mathcal R}(1-\nabla_\alpha {\mathcal R} \nabla^\alpha {\mathcal R})\right]\Big|_{r=r_H}.
\eeq
Here ${\mathcal R}=r$ denotes the areal radius, and $r_H$ the horizon location for which $f(r_H)=0$. Therefore, we obtain:
\beq
{\mathcal M}_2=\left[\frac{r}{2}\left(1+\frac{R_0R_c}{R}\right) (1-f(r)) \right]\Big|_{r=r_H}=\frac{r_H}{2}\left(1+\frac{6 R_c \beta^2 r_H}{4\Lambda r_H -6\beta}  \right)     \,.
\eeq
We remark also that any $f(R)$ gravitational theory is equivalent to the Brans-Dicke framework
\beq
S_{\text{BD}}=\int d^4 x \sqrt{-g} [ \Phi R -V(\Phi)]\,,
\eeq
where the potential $V$ of the scalar field $\phi$ is provided by \cite{bd1,bd2,bd3,oikrev}
\beq
V(\Phi)=\left( R\frac{df(R)}{dR}-f(R) \right)\Big|_{R=R(\Phi)}\,,
\eeq
with $\Phi=\frac{df(R)}{dR}$. Therefore:
\begin{eqnarray}
&& \Phi=1+\frac{R_0 R_c}{R} \quad \Rightarrow \quad R=\frac{R_0 R_c}{\Phi -1} \\
&& \Rightarrow \quad V(\Phi)=\left(R_0 R_c -R_0 \ln\frac{R}{R_c}\right)\Big|_{R=R(\Phi)} \equiv -R_0 \ln\frac{R_0}{\Phi -1}=-6 \beta^2 \ln\frac{6\beta^2 }{\Phi -1} \equiv 6\beta^2 \ln(\phi-1)\,,
\end{eqnarray}
where we have used also that potentials which differ by an additive constant are physically equivalent. Thus, in this latter formulation $\beta$ is the scaling factor of a Brans-Dicke potential. 

It should be emphasized at this point that the thermodynamical derivation of the remnant radius exhibited in \cite{ongmassive} is not affected by the underlying gravitational theory. However, the remnant entropy, if now we claim that (\ref{metric}) is a solution of (\ref{logS})	would \emph{not} be any longer (\ref{entropymassive})  but\footnote{Here we are using (\ref{eqR}).}  \cite{entropyf1,entropyf2,entropyf3,entropyf4,entropyf5}: 
\begin{eqnarray}
{\tilde S}_r &=&\frac{df(R)}{dR}\Big|_{r=R_r}  S_r= \left( 1+\frac{6 R_c \beta^2}{R}\right)\Big|_{r=R_r}  S_r= \frac{4 \Lambda R_r +6\beta(\beta R_c R_r -1)}{4\Lambda R_r -6\beta} S_r \\
&=& \begin{cases}
& \frac{ \pi [(3R_c \beta^2 +2\Lambda)\sqrt{\beta^2 +\Lambda} +\beta(3\beta^2 R_c -\Lambda)]( \beta +\sqrt{\beta^2+\Lambda})^2 }{\Lambda^3 (\beta- 2\sqrt{\beta^2+\Lambda})} \qquad {\rm if}\,\,  R_r=\frac{\beta + \sqrt{\beta^2 +\Lambda}}{\Lambda}>0, \\
&  \frac{ \pi [(3R_c \beta^2 +2\Lambda)\sqrt{\beta^2 +\Lambda} -\beta(3\beta^2 R_c -\Lambda)]( \beta -\sqrt{\beta^2+\Lambda})^2 }{\Lambda^3 (\beta + 2\sqrt{\beta^2+\Lambda})}  \qquad {\rm if}\,\,  R_r=\frac{\beta - \sqrt{\beta^2 +\Lambda}}{\Lambda},
\end{cases}     \nonumber
\end{eqnarray}
where the restrictions on the applicability of the solutions previously discussed are understood.

Moreover, the manifold (\ref{metric}) is a solution in general relativity too, given an appropriate energy-momentum tensor. However, this should not be taken as a consequence of the correspondence between the $f(R)$ gravity formulation in empty space and the general relativity formulation in which the spacetime is filled with an {\it effective nonideal}  fluid $p=\omega(\rho) \rho$ \cite{sergei}. In fact, the generalized Kiselev black hole is supported by a {\it  nonperfect} fluid whose radial and tangential pressure
\beq
p_r= -\Lambda +\frac{2\beta }{r}    \,, \qquad p_t=-\Lambda +\frac{\beta}{r}
\eeq
differ from each other \cite{anistress}. Therefore, the existence of a black hole remnant after the evaporation of (\ref{metric})  does not necessarily require the existence of a massive graviton (the thermodynamical consideration which led to the estimate of the radius of the remnant $R_r$ in \cite{ongmassive} are unaffected if we switch to the general-relativistic interpretation), because it can be due as well to the effect of a cosmic fluid with an anisotropic pressure surrounding the black hole. The Misner-Sharp mass of (\ref{metric}) in general relativity is:
\beq
\label{mrgr}
{\mathcal M}_3=\frac{r_H}{2}\,.
\eeq
Therefore when we set $r_H=R_r$ the remnant mass would be positive whenever $R_r$ is, and thus a remnant exists in both asymptotically  de Sitter and anti-de-Sitter spacetimes. We note that unlike the case of massive gravity, now we can use the plus sign in the formula for $R_r$ if $\Lambda>0$ and if $\Lambda<0$ along with $\beta<-\sqrt{|\Lambda|}$, and the minus sign only in the latter interval.

Last but not least, following \cite{grg2021} we will now construct a gravitational theory for the black hole (\ref{metric}) based on
\beq
S_{\Phi}=\frac{1}{16 \pi}\int d^4 x \sqrt{-g} {\mathcal F}(F(\Phi)R +V(\Phi)-\omega(\Phi) \nabla_\alpha \Phi \nabla^\alpha \Phi)\,,
\eeq
where $ {\mathcal F}$ is an arbitrary function of its argument, by assuming $F(\Phi)=\Phi=\frac{1}{\omega(\Phi)}$ and $V(\Phi)=0$, and restricting to the case $\Lambda=0$ for which an analytical result can be delivered. We need to solve \cite[Eq.(6)]{grg2021}:
\beq
F(\Phi)R +V(\Phi)-\omega(\Phi)\nabla_\alpha \Phi \nabla^\alpha \Phi=0\,,
\eeq
under the assumptions of staticity and  spherical symmetry for the scalar field $\Phi(r)$. Therefore:
\beq
\frac{6 \beta \Phi(r)}{r} - \frac{1}{\Phi(r)}\left(1-\frac{2M}{r}+\beta r \right)\left(\frac{d \Phi(r)}{dr}\right)^2 =0 \,,
\eeq
and we obtain:
\beq
\Phi(r)=\Phi_0 {\rm exp} \left[ \sqrt{6}  \arctan \frac{ \sqrt{\beta} \left(r+\frac{1}{2\beta}  \right)}{\sqrt{  r-2M + \beta r^2}} \right]\,, 
\eeq
where $\Phi_0$ is an arbitrary integration constant.

\subsection{Massive gravitons or anisotropic fluid?}
\label{sIVa}

The mass of the black hole remnant predicted by massive gravity (\ref{mrmassive}) would be larger than the one predicted by general relativity (\ref{mrgr}) as long as 
\begin{eqnarray}
\frac{1}{3}\left(1+\frac{\beta}{2} R_r \right) R_r > \frac{R_r}{2} \qquad \Rightarrow \qquad \beta R_r>1\,,
\end{eqnarray}
which necessarily requires a positive $\beta$, i.e. to consider the plus sign in $R_r$ together with $\Lambda>0$ (see Table \ref{table1}). Then, using $R_r=\frac{\beta + \sqrt{\beta^2 +\Lambda}}{\Lambda}$ for the radius  of the remnant, the previous condition can be recast as
\beq
\frac{\Lambda(3\beta^2 -\Lambda)}{\beta}>0,
\eeq
which delivers the constraint $\Lambda<3\beta^2 $. Therefore, for a smaller (positive) value of the cosmological constant the mass of the remnant in massive gravity can be larger than that in general relativity; this can be achieved when the pressure of the anisotropic fluid (in the Kiselev picture)  or the energy budget of the gravitons is dominating over the cosmological constant.  Recalling the proposal that remnants of primordial black holes may contribute to the dark matter budget \cite{adler1,adler2,adler3}, our result intuitively implies that  massive gravity predicts that fewer remnants are required for accounting for the same total amount of postulated dark matter. On the other hand, if we approximate the linear momentum of the remnant as $p \sim M_r$ \cite{velocity}, our result would suggest also that massive gravity may provide plausible candidates for \lq\lq warmer" dark matter than general relativity.
Furthermore, by \cite[Eq.(21)]{revpbh}
\beq
M_r \propto \frac{1}{H_{\rm form}}\,,
\eeq
where $H_{\rm form}$ is the Hubble function at the black hole remnant formation time, we can claim that for $0<\Lambda<3 \beta^2 $ the formation of the remnant  occurs at a smaller $H$, i.e. at an earlier epoch, in massive gravity than in general relativity, while for $\Lambda>3\beta^2 $ the otherwise occurs.   More quantitatively, if we write
\beq
\label{massgamma}
M_r=\frac{\gamma}{2 H_{\rm form}}\,,
\eeq
where $\gamma\simeq 0.2$ if the remnant has formed in the early universe \cite{carr}, we will obtain
\beq
\label{twocases}
\beta = \begin{cases}
	& \frac{H_{\rm form}^4 -72\gamma^2 \Lambda H_{\rm form}^2 -H_{\rm form}^2 s^{1/3} +s^{2/3}}{12 \gamma H_{\rm form} s^{1/3}} \qquad {\rm in \,\, massive\,\, gravity}, \\
	&  \frac{\Lambda \gamma^2 -4  H_{\rm form}^2}{4\gamma  H_{\rm form}}  \qquad {\rm in \,\, general \,\, relativity},
\end{cases}
\eeq
where we have defined
\beq
s:= [648\gamma^4 \Lambda^2 -180 \gamma^2 \Lambda H_{\rm form}^2 -H_{\rm form}^4+24 \gamma \sqrt {\Lambda (9 \gamma^2 \Lambda+H_{\rm form}^2)^3} ]H_{\rm form}^2  \,.
\eeq
Furthermore, by plugging (\ref{lisco}) into the previous equation in the case of general relativity, we obtain the following condition on the parameter $\beta$:
\begin{eqnarray}
&& a_2\beta^2 +a_1 \beta +a_0  =0\,, \\
&& a_1=(-3r_{\rm ISCO}^4+6r_{\rm ISCO}^2 r_p^2+r_p^4)\gamma^2+(12H_{\rm form}r_{\rm ISCO}^5-20H_{\rm form}r_{\rm ISCO}^3 r_p^2)\gamma \,, \\
&& a_2=12H_{\rm form}^2r_{\rm ISCO}^5+12r_p\gamma^2r_{\rm ISCO}^2+32H_{\rm form}\gamma r_{\rm ISCO}^4 -(20H_{\rm form}^2 r_p^2 +40H_{\rm form}\gamma r_p +9\gamma^2)r_{\rm ISCO}^3-\gamma^2 r_p^2 r_{\rm ISCO}+4\gamma^2 r_p^3 \,,\nonumber \\ &&\\
&& a_0= 4\gamma^2 r_p^2 -2(20H_{\rm form}^2r_{\rm ISCO}^3 +r_{\rm ISCO}\gamma^2)r_p+32H_{\rm form}^2 r_{\rm ISCO}^4 \,,
\end{eqnarray}
from which the entropy and physical mass of the remnant can be written in terms of its formation time, and of the size of the accretion disk and of the shadow of its progenitor.
The entropy and physical mass estimates for the case of massive gravity can be expressed in a parametric form in terms of the same quantities by following the same procedure.



Introducing the mass of the graviton $m$, we can parametrize the cosmological constant and the MOND term as \cite[Eq.(10)]{ongmassive}:
\beq
\label{parametriz}
\Lambda=3m^2(2\delta -1)\,, \qquad \beta=m^2(3\delta -1)\,.
\eeq
The relative pressure anisotropy and average pressure can be shown to be independent of the quantity\footnote{This is only a formal comparison because in the massive gravity interpretation, the black hole (\ref{metric}) is an empty space solution, but it may nevertheless be used for enlightening the geometrical meaning of $\delta$.} $m$: 
\begin{eqnarray}
\Delta&:=&\frac{\Delta p}{\bar p}=\frac{3(p_r -p_t)}{p_r +2p_t}=\frac{3(1-3\delta )}{6(3 r-2)\delta-9r +4} \,,\\
w&:=&\frac{\bar p}{\rho}=\frac{p_r +2p_t}{3 \rho}=\frac{6(3r-2)\delta-9r+4}{18(r-1)\delta+3(2-3r)}\,.
\end{eqnarray}
Applying the parametrization (\ref{parametriz}) to the massive gravity case in (\ref{twocases}), and considering a series expansion for small $m\simeq0$, we obtain the second-order equation
\beq
\frac{H_{\rm form}}{12 \gamma}+\frac{(3\delta-1)H_{\rm form}+2(1-2\delta)\gamma}{H_{\rm form}}m^2 \approx 0\,,
\eeq
from which we can then eliminate the parameter $\delta$ via the cosmological constant getting
\beq
8 \Lambda\gamma^2 - 6H_{\rm form}(m^2+\Lambda)\gamma+H_{\rm form}^2 \approx 0\,.
\eeq
Therefore the mass of the graviton in terms of the cosmological constant and of the formation epoch of the remnant is
\beq
\label{massgrav}
m \approx \sqrt{\frac{8\Lambda\gamma^2-6H_{\rm form}\Lambda\gamma-H_{\rm form}^2}{6 \gamma H_{\rm form}}} \,.
\eeq
This result is well defined if $2\gamma(4\gamma -3 H_{\rm form})\Lambda>H_{\rm form}^2$, i.e. $H_{\rm form}>4 \gamma/3$ assuming $\Lambda<0$ which constitutes a lower bound on the epoch of formation of the remnant. By using  (\ref{massgamma}), an upper bound is set on the mass of the remnant as $M_r<3/8$. Restoring International System units by multiplying by $c^2/G$, we get the bound on the remnant mass of $M_r<0.5 \cdot 10^{27}$ kg, which is 1/4000 solar masses. We remark that we have never used any specific value for $\gamma$ in the steps leading to this estimate.

On the other hand, if we specifically impose the linearized limit of our massive gravity paradigm to admit  two degrees of freedom, the mass of the graviton should be $m^2=-2 \Lambda/3$ \cite{novello}. Then, by (\ref{massgrav}) the formation time of the remnant predicted by massive gravity is
\beq
H_{\rm form} \approx (-\Lambda+\sqrt{\Lambda^2+8\Lambda})\gamma  \,
\eeq
which is well-defined if $\Lambda<-8$. In this case by using (\ref{massgamma}) the mass of the remnant would be given by $M_r \approx \frac{1}{2(-\Lambda+\sqrt{\Lambda^2+8\Lambda})}$.

\section{Conclusion}
\label{sV}

In this paper, we have investigated the role that a linear term in the redshift function of a black hole is playing from both the geometrical and astrophysical perspectives. The strength of this term is quantified via the parameter $\beta$, and it arises in few different theories (general relativity, Weyl, massive and $f(R)$ gravity). Physically the linear term has different origin.
In the general relativity framework, in which our black hole spacetime is known as the {\it generalized Kiselev solution}, the black hole is surrounded by an imperfect fluid, and the linear term is due to an uneven balance between radial and transversal pressure with $p_r -p_t=\beta$ at fixed radius.  In the massive gravity interpretation of the same black hole manifold such behavior would be caused by an higher value of the graviton mass. However,  observables based only on the metric and its derivatives cannot distinguish between these theories. 

On the other hand, the metric degeneracy is broken when thermodynamics is taken into account. Specifically, when interpreted in light of these physically different theories the
black hole entropy may be different. Furthermore the Hawking evaporation process of the black hole would deliver remnants with different values of their physical masses with direct consequences on their viability as dark matter candidates.

\begin{acknowledgments}
DG is a member of the GNFM working group of the Italian INDAM. YCO thanks the National Natural Science Foundation of China (No.11922508) for funding support.  B.W. was also partially supported by NNSFC under grant No. 12075202. 
\end{acknowledgments}

{}

\end{document}